\documentclass[aps,prd,groupedaddress]{revtex4}

\usepackage{graphics}
\usepackage{rotating}

\begin{document}

\title{$K\bar K$ molecules with momentum--dependent interactions } 

\author{R. H. Lemmer}

\affiliation{School of Physics, University of the Witwatersrand, Johannesburg,
Private Bag 3, WITS 2050, South Africa}

\date{\today}

\begin{abstract}
   It is shown that the  momentum--dependent kaon--antikaon interactions
   generated via vector
   meson exchange from the standard $SU_V(3)\times SU_A(3)$ interaction
   Lagrangian lead to a non--local potential in coordinate space that can be
   incorporated without approximation into a non--relativistic version of the
   Bethe--Salpeter wave equation containing a radial--dependent
   effective kaon mass appearing in a fully symmetrized kinetic energy operator,
   in addition to a local potential.  Estimates of the mass and decay widths
   of $f_0(980)$ and  $a_0(980)$, considered as $K\bar K$ molecules of isospin $0$ and
   $1$, as well as
   for $K^+K^-$ atomic bound states (kaonium) are presented, and compared with previous
   studies of a similar nature.  It is argued that without a better knowledge
   of hadronic form factors it is not possible to distinguish between the
   molecular versus elementary particle models for the structure of the light scalar mesons.


\vspace{10cm}
\noindent

\vspace{1cm}
Corresponding author: R H Lemmer

Electronic address: rh$_-$lemmer@mweb.co.za

\end{abstract}
\maketitle

\newpage
 \section{Introduction}
  
  The properties of  multi quark--antiquark interacting systems as
  investigated by Weinstein and
  Isgur \cite{WIsgur82} pointed to a possible bound kaon--antikaon
  molecular state
   structure for the light  scalar mesons $f_0(980)$  and $a_0(980$).
   Alternative proposals include, in addition to the molecular
   picture \cite{WIsgur82,TB85,SU390,JAO03,KLS04,TBR08},
    a $q\bar q$ state \cite{LM00} or 
  a $q^2\bar q^2$ state \cite{NNA89}.
   However,
   the large $2\pi$ annihilation width for
   $q\bar q\to \pi\pi\sim 500$ MeV from
   flux tube--breaking \cite{RKNI87} to $\sim 600 $ MeV using current algebra \cite{RT02},
   disqualifies a  light quark--antiquark configuration when compared with
   a typical experimental width  of  $\sim 50$ MeV  for
   the $f_0(980)\to\pi\pi$ decay \cite{PDG08}.  The inclusion of
   strange quarks  changes
   this picture \cite{VD96,MDS04}.
   For example, it is shown in \cite{MDS04}
   that if the $f_0(980)$  is identified with the
   (almost) pure $s\bar s$ member of the  nonet within a quark--level
   linear $\sigma$ model  framework, one can obtain  reasonable values for the mass as well
   as $\pi\pi$ and $\gamma\gamma$ decay widths.
   So the question of a  molecular versus  a $q\bar q$ elementary particle
   structure, or perhaps some combination
   of these, for the light scalars remains an open one \cite{PEN07}.

    In the following we re--investigate the
      molecular  state option
    using the non--local extension of the vector meson
    exchange potentials derived
    in \cite{KLS04} for  kaon--antikaon bound states and their decay modes.
    It is shown that if one
     adheres strictly to
    the approximations required \cite{LL74,WLFS01} for reducing the instantaneous version of the
    Bethe--Salpeter
    equation to a non--relativistic wave equation, especially as regards the off--shell
    nature of the relative four--momentum of the boson pair undergoing binding, a quantitatively similar
    picture to that given by a purely local approximation to the  potential
    emerges for kaon--antikaon bound states.
    In particular, the non--binding states reported in a recent analysis \cite{YHPB06} of the
    same problem  now no longer occur.

     The paper is arranged as follows: Section II  discusses possible 
     kaon--antikaon bound states 
    of the coordinate space
  version of the
  non--relativistic Bethe--Salpeter equation for non--local interactions
    as generated by vector meson exchange from a $SU_V(3)\times
   SU_A(3)$ invariant interaction Lagrangian. In Section III  we
   calculate the dominant decay widths of such bound $K\bar K$ pairs
   of good isospin for annihilation into either $\pi\pi$ or $\pi^0\eta$,
  while in Section IV the low energy $K\bar K$ scattering
  parameters in the presence of annihilation
  are found. Finally, Section V is devoted to  a brief discussion of the
  energy shifts and decay widths to be expected
  for the associated   mesonic atom $K^+K^-$ (kaonium)
   in context of the non--local interactions introduced in this paper.
  A summary and conclusions follow in Section VI.

 \section{ $K\bar K$ bound states}
 \subsection{Non--relativistic Bethe--Salpeter equation}

   As before \cite{KLS04}
   we use the non--relativistic version of the Bethe--Salpeter (BS)
  equation  in the instantaneous approximation \cite{LL74,WLFS01},
  described further below,
  to study the mass and  decay widths  of
  the $f_0(980)$ and $a_0(980)$ scalar mesons, considered
  as  weakly bound kaon--antikaon pairs of isospin zero or one interacting
  via vector meson exchange. In momentum space this equation reads
  \begin{eqnarray}
  \Big(\frac{{\bf p'\:^2}}{M_K}+2M_K-P_0\Big)\phi({\bf p'})=
   \frac{1}{4M_K^2}\int \frac{d^3p}{(2\pi)^3}
  \hat\Gamma({\bf p',p};P_0)\phi({\bf p})
  \label{e:BSm}                         
  \end{eqnarray}
  where $\phi({\bf p})$ is the wave function of the interacting pair in
  momentum space; $M_K \approx 496$ MeV is the average kaon mass. The  eigenvalue $P_0=2M_K+E$ is the
  as yet unknown total mass of the interacting system of binding energy
  $E=-\epsilon<0$.

   Apart from introducing non--relativistic particle propagators, the
    other essential simplification used in deriving Eq.~(\ref{e:BSm}) is 
    the assumption that the interaction is instantaneous.
    This is implemented
    by suppressing the time components of the relative four--momenta
 $p=(p_1-p_2)/2\to [0,{\bf p}]$ and $p'=(p'_1-p'_2)/2\to [0,{\bf p'}]$ in
   the irreducible four--point vertex, or  transition amplitude $(p'_1,p'_2|\hat\Gamma|p_1,p_2)$
    of the original
   BS equation \cite{LL74}. For a bound state the four--momenta
    that enter and leave this vertex are all off their mass shell. In the instantaneous
    approximation in particular this means that  
   $(p_1,p_2)=[\frac{1}{2}P_0,{\bf \pm p}]$ and $(p'_1,p'_2)=[\frac{1}{2}P_0,{\bf \pm p'}]$
   in the center of mass (c.m.) system,
   where $p_1+p_2=
    p'_1+p'_2=[P_0,0]$ still guarantees total energy--momentum conservation for
    arbitrary values of ${\bf p}$ and  ${\bf p'}$. These approximations 
    result in the transition amplitude
      \begin{eqnarray}
   (p'_1,p'_2|\hat\Gamma|p_1,p_2)=(p',P_0|\hat\Gamma|p,P_0)\approx
   \hat\Gamma({\bf p',p};P_0)
  \label{e:gamma}
  \end{eqnarray}
  which appears in Eq.~(\ref{e:BSm})
   that only depends on the unconstrained three--momentum of the incoming and
   outgoing kaons in addition
   to the total energy, and legitimizes  integration over these independent
   momentum variables.

   An expression
   for $\hat\Gamma({\bf p',p};P_0)$ for
   $K\bar K$ scattering via $t$--channel vector meson exchange has been derived
     in \cite{SU390,KLS04} and also \cite{YHPB06} from
     a standard $SU_V(3)\times
     SU_A(3)$ invariant interaction Lagrangian \cite{DESWART63} to which we
     refer for further details.
     One finds in this case that
     $(p'_1,p'_2|\hat\Gamma|p_1,p_2)$ 
      is given to lowest order coupling by
     \begin{eqnarray}
     (p'_1,p'_2|\hat\Gamma|p_1,p_2)= -g^2_iC_in_s 
     \Big[\frac{(p_1+p'_1)\cdot(p_2+p'_2)
     +(p^2_1-p'^{\;2}_1)(p^2_2-p'^{\;2}_2)/M^2_i}{(p_1-p'_1)^2-M^2_i}\Big]
    \label{e:relvertex}
     \end{eqnarray}
   that reduces to the
  approximate
  transition amplitude introduced in Eq.~(\ref{e:gamma}),
                         
    \begin{eqnarray}
 (p'_1,p_2'|\hat \Gamma_i|p_1,p_2)&\approx& \hat\Gamma_i({\bf p',p};P_0)
  =g^2_{i} C_in_s\Big[\frac{P_0^2+({\bf p\;'+\bf p})^2
 +({\bf p}\;'^2-{\bf  p}^{\;2})^2/M^2_i}{({\bf p\;'-\bf p})^2+M^2_i}\Big]
 \label{e:vertex}
  \end{eqnarray}
  for the  values of the off--shell
  four--momenta prescribed above. This vertex structure also holds true for $K\bar K\to \pi\pi$
  where ${\bf p}'$ now refers to the outgoing pion momentum,
  but not for $K\bar K\to \pi^0\eta$ without modification, see below Eq.~(\ref{e:gamfnl}).

 The  coupling constants  $g^2_i$ are all related to the
  the $\rho\pi\pi$ coupling constant $g_{\rho\pi\pi}$ 
  by the $SU(3)$ symmetry \cite{DESWART63}:
 $g_i^2=(\frac{1}{4},\frac{1}{4},\frac{1}{2})g^2_{\rho\pi\pi}$
   for $i=(\rho,\omega,\phi)$ exchange in $K\bar K\to K\bar K$,
  or $g^2_i= (g^2_{\pi K K^*},g_{\pi K K^*} g_{\eta K K^*})=(\frac{1}{4},
  \frac{1}{4}\sqrt 3)g^2_{\rho\pi\pi}$ for $K^*$ exchange leading
  to $K\bar K\to \pi\pi$ or $\pi^0\eta$ transitions. $M_i$ is the mass of the
  exchanged meson in each case,
and $\alpha_s=g^2_{\rho\pi\pi}/4\pi\approx 2.9$ is known from
 the KSRF relation \cite{KSRF}. The $C_i= [3,1,1]$ or $[-1,1,1]$
 are isospin factors \cite{SU390} for 
  $(\rho,\omega,\phi)$ exchange between a $(K\bar K)_I$ pair coupled to $I=0$
  or $1$, or
$C_i=[-\sqrt 6,-\sqrt 2]$ for the $I=0$ and $1$ 
  transitions $(K\bar K)_{0,1}\to (\pi\pi)_0$ or  $(\pi\eta)_1$
  via $K^*$ exchange;
  $(K\bar K)_1\to (\pi\pi)_1$  is forbidden
 by G parity conservation;
  $n_s$ is a symmetry factor
 $[\frac {1}{\sqrt 2},1]$ for identical (non--identical) bosons
 in the final state.

 As already remarked in \cite{SU390}, one can check
 the relations  between the various vector--pseudoscalar coupling constants
  given above empirically by appealing to the experimental decay widths of those vector
  mesons $i=(\rho,K^*,\phi)$
  that are unstable with respect to  decay into
  $\pi\pi$, $K\pi$ and $K\bar K$.  An elementary calculation shows that a vector meson of mass
  $M_i$ at rest has a decay width of
  \begin{eqnarray}
  \Gamma_i=n_s^2f_i\Big[\frac{2}{3}\frac{p^3_i}{M^2_i}\frac{g^2_i}{4\pi}\Big]
  \label{e:cc}
  \end{eqnarray}
  in the scheme of \cite{SU390} to first order in the relevant coupling constant $g^2_i$, where $p_i$ is the
  common c.m. momentum magnitude of either decay particle in the final state, and $f_i=(2,3,2)$ an
  isospin factor for the three  channels in question.
  The symmetry factor $n_s$ has been defined above. Inserting the known \cite{PDG08}
  masses and full or partial decay widths  of $(\rho,K^*, \phi)$ as appropriate
  into   Eq.~(\ref{e:cc}),  one finds
  $(g^2_{\rho\pi\pi}:g^2_{\pi K K^*}:g^2_{\phi K\bar K})\sim (1:(0.54)^2:(0.77)^2$)
  as compared with  $(1:(1/2)^2:(1\sqrt{2})^2)$ given above, where $g^2_{\rho\pi\pi}/4\pi=2.92$ that
  coincides almost exactly
  with the KSRF value. The remaining empirical ratios  hold to within $\sim 8\%$ 
  of the ``ideal'' mixing version of the  $SU(3)$ relations between the coupling constants.

  The
  expression for the two--particle vertex given by Eq.~(\ref{e:vertex})
  regards the basic  vertices as point--like. Upon introducing a
  form--factor  depending on the three--momentum transfer ${\bf q=p'-p}$ of the
  form \cite{SU390}
  \begin{eqnarray}
  F_i({\bf q}^2)=\Big(\frac{ 2\Lambda^2-M_i^2}{2\Lambda^2+{\bf q^2}}\Big)^2
  \label{e:ff}
  \end{eqnarray}
  at each  of these vertices with a common high--momentum
  cutoff $\sim \sqrt{2}\Lambda$, 
  the vertex in Eq.~(\ref{e:BSm}) is altered to read
 $\hat \Gamma({\bf p',p};P_0)=\sum_{i} \hat \Gamma_i({\bf p',p};P_0)F^2_i({\bf q}^2)$.
 The use of a common $\Lambda$ is simply a working
 assumption;  the cutoffs could in principle be different for different
 vertices \cite{BUGG04}.

    \subsection{Non--local interaction potential in coordinate space}
  The  vertex in Eq.~(\ref{e:vertex}) was approximated further in \cite{KLS04}
  by neglecting the small binding energy $\epsilon\sim\frac{1}{50}M_K$
  contribution
 by taking $P_0\approx 2M_K$ and  ignoring the three--momentum dependence
  in the numerator entirely.
  Then the
  components of the $K \bar K$ potential on the right hand side of
  Eq.~(\ref{e:BSm}) reduce to
  \begin{eqnarray}
  \frac{1}{4M_K^2}\hat\Gamma_i({\bf p',p};P_0)F^2_i({\bf q^2})
  = g^2_{i}C_i U_i({\bf q}),\quad U_i({\bf q})=\frac{F^2_i({\bf q^2})}{M_i^2+{\bf q^2}}
  \label{e:local}
   \end{eqnarray}
   that is only a function of the momentum transfer ${\bf q^2}$.
   This  leads to the following set of
   local potentials in coordinate space for each exchanged meson  $M_i$
     that are known
   in closed form   from the fourier transform of $U_i({\bf q})$,
    \begin{eqnarray}
    &&V_i(r)= - C_ig^2_iU_i(r),\quad
    U_i(r)=\frac{1}{4\pi}\Big\{\frac{e^{-M_i r}}{r}-\frac{e^{-\sqrt 2 \Lambda r}}{r}
     \sum_{n=0}^3 C^{(0)}_n(\sqrt 2\Lambda r)^n\Big\}.
   \label{e:basicpot}
   \end{eqnarray}
    The $ C^{(0)}_n$'s are polynomials
     $ C^{(0)}_0=1,\;  C^{(0)}_1=\frac{1}{16}(11-4c_i^2+c_i^4)(1-c_i^2),\;
      C^{(0)}_2 =\frac{1}{16}(3-c_i^2)(1-c_i^2)^2$ and
     $ C^{(0)}_3=\frac{1}{48}(1-c_i^2)^3$ in the variable 
     $c_i=M_i/\sqrt 2 \Lambda$.
     The partial potential $V_i(r)$ has a simple power series expansion
   in $r$ about the origin when the form factor is included,
    \begin{eqnarray}
    V_i(r)=-\sqrt 2 \Lambda\alpha_iC_i(1-c_i)^4\Big\{\frac{1}{16}(5+4c_i+c^2_i)-
   \frac{1}{96}(1+4c_i+c^2_i)(\sqrt 2 \Lambda r)^2+\cdots\Big\}
   \label{e:expand}
   \end{eqnarray}
 with $\alpha_i=g^2_i/4\pi$,  
   so that the
    $V_i(r)$ as well as its space
   derivatives  are all well--behaved at $r=0$ for finite $\Lambda$.

   In general the interaction vertex in Eq.~(\ref{e:BSm}) leads to
   a non--local potential in coordinate space. However, since the numerator
   in the special case of Eq.~(\ref{e:vertex}) is a  polynomial in the
   three--momentum variables, one can incorporate them in coordinate
   space in a revised
   potential  that  also involves space derivatives of $V_i(r)$  coming from the
   replacements ${\bf p}$ or ${\bf q}\to -i\nabla$ in
   $\hat\Gamma_i({\bf p+q,p};P_0)$. Again setting 
    $P_0\approx 2M_K$, the  $K\bar K$ non--local
    interaction potential operator $V_{K\bar K}$ becomes 
    \begin{eqnarray}
    V_{K\bar K}\psi(r)&=&V_3(r)\psi(r)
    -\frac{1}{4M_K^2}\Big[V_2(r)\nabla^2+2\nabla \cdot V_2(r) \nabla +
    \nabla^2 V_2(r)\Big]\psi(r)
    \label{e:vees}
    \end{eqnarray}
  when operating on the s--state wave function $\psi(r)$, 
  the three--dimensional fourier transform of the
  spherically symmetric momentum space wave function $\phi(p)$ of
  Eq.~(\ref{e:BSm}).

    The potentials  $V_3(r),V_2(r),V_1(r)$ are given by the following combinations
     (the primes indicate derivatives with respect
    to $r$)
   \begin{eqnarray}
     &&V_3(r)= V_1(r)+\frac{1}{2M_K^2}\frac{1}{r}(V'_1(r)-V'_2(r))
    \nonumber
    \\
    &&V_2(r)= \sum_i\Big(V_i(r)-\frac{V_i''(r)}{M^2_i}\Big)
    \nonumber
    \\
     &&V_1(r)=\sum_i V_i(r).
     \label{e:pots12}
     \end{eqnarray}
  Then the coordinate
  space version of Eq.~(\ref{e:BSm}) with the full momentum structure of
  Eq.~(\ref{e:vertex}) included can be written as
  
   \begin{eqnarray}
   -\frac{1}{4}\Big[\frac{1}{M^*_K(r)}\nabla^2+2\nabla \cdot
   \frac{1}{M^*_K(r)} \nabla+\nabla^2\frac{1}{M^*_K(r)}\Big]\psi(r)+
    V_3(r)\psi(r)=E\psi(r)
  \label{e:waveeqn}
  \end{eqnarray}
 containing a spatially--dependent  ``effective'' kaon mass
 \begin{eqnarray}
  M^*_K(r)=\gamma^{-1}(r)M_K,\quad \gamma(r)=(1+\frac{V_2(r)}{M_K})
  \label{e:effmass}
  \end{eqnarray}
  that enters into the equation via  a fully symmetrized  kinetic energy  operator.

  The quartic terms in the vertex
  Eq.~(\ref{e:vertex}) only contribute to
  off--shell scattering; they vanish on--shell when ${\bf p}'^2={\bf p^2}$.
  This suggests that their contribution to the interaction potential may be small.
  If they are omitted entirely,
   both  $ V_3(r)$ and $V_2(r)$ reduce to $ V_1(r) $ and  the structure
  of Eq.~(\ref{e:waveeqn}) becomes
  identical in form with that of a wave equation derived in \cite {WFRL57}
  that has been used quite generally for describing non--local
  effects on particle motion in nuclei \cite{PFRL96}.

The hermitian structure of the symmetrized kinetic energy operator
assures that the eigenstates of  Eq.~(\ref{e:waveeqn})
 remain orthogonal, and also that the
 continuity equation
for its time--dependent version  continues to hold
with a revised probability current density 
${{\bf j}(r,t)}=1/iM^*_K(r)[\psi^*\nabla \psi-\psi\nabla\psi^*]$. Thus
the normalization of $\psi(r,t)$ also remains independent of time.

  \subsection{Numerical results}
    We next investigate  possible bound $s$--state solutions
    of Eq.~(\ref{e:waveeqn}). It is convenient to set
    $\psi(r)=\psi(0)u(r)/r$ with $u(r)/r\to 1$ as $r\to 0$.
     The value of $\psi(0)$ is  fixed by normalization.
     Then Eq.~(\ref{e:waveeqn}) 
    simplifies to 
   \begin{eqnarray}
     \gamma(r) u''+\gamma'(r)u'+M_K(E-V_0)u=0,
   \label{e:radialeqn}
   \end{eqnarray}
    where
   \begin{eqnarray}
   V_0(r)=V_1(r)+\frac{1}{2M_K^2}\frac{1}{r}V'_1(r) -\frac{1}{4M_K^2}V''_2(r).
   \label{e:vezero}
   \end{eqnarray}
   We  solve this equation numerically after using the partial potentials
   given by Eq.~(\ref{e:basicpot}) to construct the required interactions.
   For $I=0$ one has 
   \begin{eqnarray}
    V_1(r)= -g^2_{\rho\pi\pi}\Big(\frac{3}{4}U_\rho(r)+\frac{1}{4}U_\omega(r)
    +\frac{1}{2}U_\phi(r)\Big)
    \end{eqnarray}
    and
    \begin{eqnarray}
    V_2(r)=-g^2_{\rho\pi\pi}\Big[\frac{3}{4}
     (U_\rho(r)-\frac{1}{M_\rho^2}U''_\rho(r))+
     \frac{1}{4}
     (U_\omega(r)-\frac{1}{M_\omega^2}U''_\omega(r))+
     \frac{1}{2}
     (U_\phi-\frac{1}{M_\phi^2}U''_\phi(r))\Big]
    \end{eqnarray}
    from which $V_0(r)$ can be obtained. For $I=1$ identical expressions
    hold with the coefficient $3/4$ of the $\rho$ meson exchange contribution
     replaced by $-1/4$.   

    Since  known physical masses of  exchanged mesons
    $(M_\rho, M_\omega,M_\phi)=(768,783,1019)$ MeV
    and their coupling constants given above enter the calculations,
     the only free parameter  is the cutoff $\Lambda$. 
     The possible choices of
    $\Lambda$ are then restricted further
   if one in addition 
   requires that $\gamma(0)>0$, in order to ensure that $M^*_K(r)$ 
   remains positive in the interaction zone \footnote{Without this restriction
   a physical problem arises because Eqs.~(\ref{e:waveeqn}) or (\ref{e:radialeqn})
   then develop a regular singular point  at $r=r_s >0$ where $\gamma(r_s)=0$.
   There is no difficulty in solving the resulting differential equation and continuing
   the solution through $r_s$, where $u(r_s)$ is finite but has an infinite first
   derivative there. However such solutions lead to a probability current density that
   diverges at the singularity.}.
   This problem has already
   been encountered in  \cite{YHPB06},
   where it is concluded that for cutoffs
   that ensure (their equivalent of) $\gamma(0)>0$ give non--local potentials                
   that do not bind the $K\bar K$ pair. However these
   potentials  have been based on an approximated vertex that corresponds to 
   $\hat\Gamma_i({\bf p',p};P_0)$ of Eq.~(\ref{e:vertex}) without the quartic
   contribution and $P^2_0$ replaced by the combination $2(E^2_{p'}+E^2_{p})$ of
   kaon c.m. energies.
   The first approximation neglects a small correction term, but the second one  
   excludes  off--shell  kaons, $p\:'\neq p$, by total energy conservation, 
   contrary to what is expected for bound states.
   Thus these potentials may not correctly reflect the underlying physics for such systems \cite{LL74}.
                              
\begin{figure}
\rotatebox{0}{\includegraphics{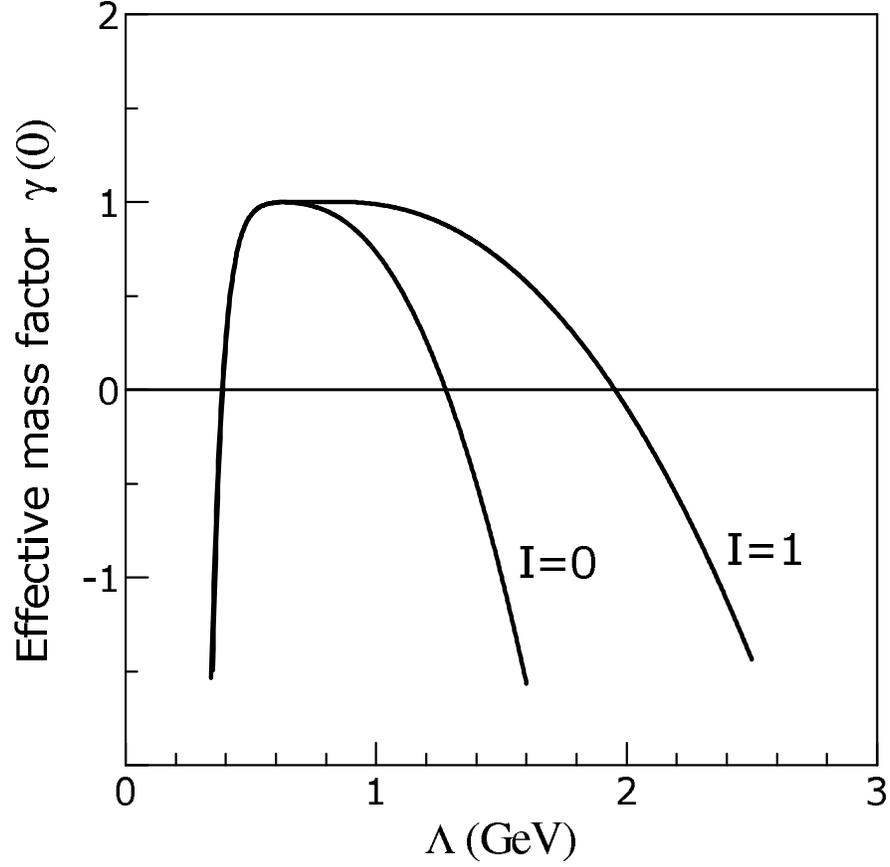}}
\caption{\label{f:fig1}
Critical values of $\gamma(0)$, the inverse effective mass ratio  versus the cutoff
$\Lambda$ for isospin $I=0$ and $I=1$. Both curves remain positive in the
common interval $\sim 0.39\alt \Lambda \alt 1.28$ GeV. }

\end{figure}

  This is not the case for the  potential operator $V_{K\bar K}$ of  Eq.~(\ref{e:vees}).
  Using Eqs.~(\ref{e:expand}) and (\ref{e:pots12}) one readily finds that
 \begin{eqnarray}
 \gamma(0)=1 -\frac{\sqrt 2 \Lambda}{M_K}\sum_i \alpha_iC_i(1-c_i)^4\Big\{\frac{1}{16}(5+4c_i+c^2_i)
 + \frac{1}{48}(1+4c_i^{-1}+c_i^{-2})\Big\}.
 \label{e:V2zero}                           
 \end{eqnarray}
 For the known interaction
 strength of $\alpha_s=g^2_{\rho\pi\pi}/4\pi=2.9$, the factor $\gamma(0)$ 
 stays positive in both isospin channels if $\Lambda$ lies in the interval $390\alt\Lambda
 \alt 1280$ MeV, see Fig.~\ref{f:fig1}.
 However whether or not the resulting non--local potentials
 bind the $K\bar K$    pair is still sensitive to the actual choice  of cutoff.
    Table~\ref{t:table1} lists calculated values of binding energies,
    masses and decay
    half--widths (see next section for the latter)
    for the  subset of cutoff values $\Lambda =(415,\; 420,\; 425)$ MeV
    confined to a relatively small window within
    this interval.
    These cutoffs have been chosen so as
    to  bracket the spread in experimental \cite{PDG08} binding
    energies  
     $ 11.3\pm 10$ MeV for the $f_0(980)$ 
      meson  considered
     as an interacting $K\bar K$ pair. As a byproduct the $a_0(980)$ 
     then also appears as  a  bound pair state in the isovector channel of
     these potentials.

     A much stronger purely
     local potential \cite{KLS04} of depth $\sim 4$ GeV was required to give the same
     order of binding for $I=0$ (and none at all
     for $I=1$).
    The reason for this is clear from Table~\ref{t:table2} which summarizes 
    results for the isoscalar potential. We see that the much
      shallower non--local potential $\sim 0.2$ GeV is counter--balanced
     by a  significant suppression  of the kinetic energy through an increased
     effective kaon mass of $M^*_K(0)\approx 2M_K$ at the origin that favors binding.
     One notes  that  cutoff momenta
     $q_{max}= \sqrt{2}\Lambda\sim M_\eta$, of order
     of the $\eta$ meson
     mass, that also sets the mass scale of the pseudo--scalar meson octet, are required
     to regulate the
      behavior of the bare vertex in 
       Eq.~(\ref{e:vertex}) at large momentum transfer. Otherwise this vertex  either
       tends to a constant or diverges  depending on whether the fourth order
       contribution is  dropped or kept.
     These  cutoffs are typically an order of  magnitude  smaller than those 
      for a local potential like that in Eq.~(\ref{e:basicpot})
     to give the same order of binding. We also comment that the results in Table~\ref{t:table1}
     remain qualitatively unchanged by dropping the quartic momentum contributions
     altogether and setting $V_3(r)=V_2(r)=V_1(r)$ in Eqs.~(\ref{e:waveeqn})
     and (\ref{e:effmass}).

  \section{decay of $K\bar K$  bound states}
\begin{figure}
\rotatebox{0}{\includegraphics{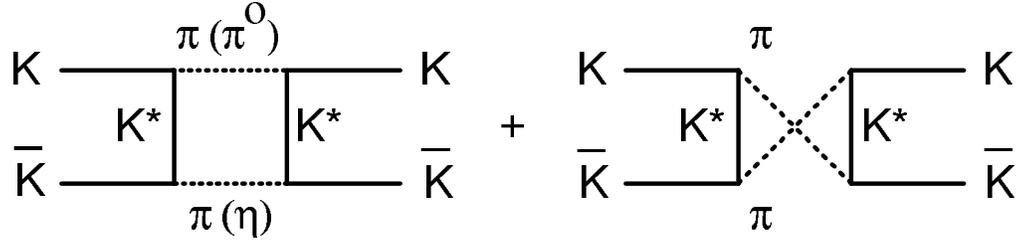}}
\caption{\label{f:fig2}
  The direct and exchange diagrams that contribute to the additional four--point vertex,
  or box
  diagram $\hat\Gamma_{box}({\bf p}',{\bf p};P_0)$,
  in the BS equation for $K\bar K$ scattering that arises from $t$-channel $K^*(892)$
  exchange between the kaons, leading
  to $\pi\pi$ isoscalar or $\pi^0\eta$ isovector intermediate states.
  In the latter case only the first diagram is present.}
\end{figure}

   The annihilation of  bound $K\bar K$ pairs into mesons or photons
   has been discussed in \cite{KLS04} and \cite{RHL07,HANR07} in the context of the BS equation. The                     
  decay of the  bound  state is governed by the imaginary part of the
  additional box diagram $\hat\Gamma_{box}({\bf p'},{\bf p};P_0)$ contribution
  to the  
   BS equation vertex where a $K^*$ vector meson is exchanged in the $t$--channel, see
   Fig.~\ref{f:fig2}.
   Then the eigenvalue $P_0$ in Eq.~(\ref{e:BSm}) acquires an imaginary part,
   $P_0\to P_0-i\Gamma/2$.   The imaginary part of the box diagram can be
   retrieved
   directly by applying 
   the Cutkosky rules \cite{LL74} to cut the intermediate meson loops in
   Fig.~\ref{f:fig2}. The decay widths $\Gamma$  are then obtained
   perturbatively by averaging over the bound state solutions of
   Eq.~(\ref{e:BSm}) in the absence of annihilation.

     For the  $I=0$, $K\bar K\to \pi\pi$ decay this  width
     reads
  \begin{eqnarray}
  \Gamma_{\pi\pi}=
   \frac{1}{32\pi^2}\frac{p_\pi}{M_K}\frac{1}{4M^2_K}\int d\Omega_\pi
   |M_{B}({ p}_\pi)|^2                          
   \label{e:gampipi}
  \end{eqnarray}
  where the bound to free transition amplitude $M_B(p_\pi)$ can 
   be written as either a momentum or a coordinate space integral,
   \begin{eqnarray}
   M_{B}({ p}_\pi)=\int\frac{d^3p}{(2\pi)^3}\phi({\bf p})
   M_{\pi\pi}({\bf p}_\pi,{\bf p})=\int d^3 r\psi({\bf r})
   M_{\pi\pi}({\bf p}_\pi,{\bf r})
   \label{e:MPI}
   \end{eqnarray}
   by  fourier transforming $\phi({\bf p})$ and
   $M_{\pi\pi}({\bf p}_\pi,{\bf p})$. 
   The angular integral  $d\Omega_{\pi}$ in Eq.~(\ref{e:gampipi}) runs over the full
  solid angle of one of the pions due to wave function symmetrization \cite{SU390}.

    The amplitude $ M_{\pi\pi}({\bf p}_\pi,{\bf p})$ refers to
    $K\bar K$ annihilation  
    where the pions are  on--shell with the magnitude of their
    common momentum fixed by total energy conservation at
    $p_\pi = (\frac{1}{4}P^2_0-m^2_\pi)^{1/2}$;
    the kaons remain off--shell as before.
    $ M_{\pi\pi}({\bf p}_\pi,{\bf p})$ is then given by
    $\hat\Gamma_i({\bf p}_\pi,{\bf p};P_0)
    + \hat\Gamma_i(-{\bf p}_\pi,{\bf p};P_0)$
     from Eq.~(\ref{e:vertex}) with
     $g^2_iC_in_s= -\sqrt 3 g^2_{\pi K K^*}$ and $M_i=M_{K^*}$.
       The $ {\bf p}_\pi\to -{\bf p}_\pi$
     crossed contribution comes about since the pions are identical bosons
     in the symmetrized space $\times$ isospin basis \cite{SU390}.
     Performing the fourier transform of this sum gives the transition
     operator  in coordinate space as                         
   \begin{eqnarray}
    M_{\pi\pi}({\bf p}_\pi,{\bf r})=-2\sqrt 3g^2_{\pi K K^*}e^{-{i{\bf p}_\pi\cdot r}}
    \bigg\{\delta^3({\bf r})+\Big[P^2_0+4p^2_\pi-M^2_{K^*}
    +4i({\bf p}_\pi\cdot{\bf\nabla})\Big]\frac{e^{-M^* r}}{4\pi r}\bigg\}
 \label{e:transr}
 \end{eqnarray}
  after discarding
  the quartic terms of order
  ${\cal O}(M^2_K/4 M^2_{K^*})\ll 1$.
  The delta function contribution
 arises from the
  quadratic dependence on the momentum transfer ${\bf q}={\bf p}-{\bf p}_\pi$
  in the numerator of
  Eq.~(\ref{e:vertex}) that introduces the Laplacian $\nabla^2$ operating on
   the  Yukawa--like potential in coordinate space
  produced by the $K^*$ meson exchange.
  We insert this  result into the second form of Eq.~(\ref{e:MPI}) with
  $\psi({\bf r})=\psi(0)u(r)/r$ to
 find

  \begin{eqnarray}
   M_B(p_\pi)=- 8\pi\alpha_{\pi\pi}\psi(0) R(p_\pi)
  \label{e:MBfinal}
   \end{eqnarray}
  where $\alpha_{\pi\pi}$ is an effective coupling constant at zero kaon
  momentum   
 \begin{eqnarray}
 \alpha_{\pi\pi}= -\frac{\Gamma_i({\bf p}_\pi,0;P_0)}{4\pi}=\frac{1}{4}
 \sqrt 3\;\alpha_s 
 \Big[\frac{P_0^2+ p^2_\pi}{M^2_{K^*}+ p^2_\pi} \Big] \approx 1.489
 \label{e:apipi}
 \end{eqnarray}
 for $\alpha_s=2.9$ and $P_0\approx 2M_K$.
This determines the annihilation cross section for free  $K\bar K$
  pairs into two pions at low momentum as
  \begin{eqnarray}
  \sigma_0=\frac{\pi\alpha^2_{\pi\pi}}{M^2_K}\frac{p_\pi}{p}.
  \label{e:sigzero}
   \end{eqnarray}
   The factor
  $R$ is given by 
   \begin{eqnarray} 
 &&R(p_\pi)=                                              
   \Big[\frac{P_0^2+p^2_\pi}{M^2_{K^*}+p^2_\pi}\Big]^{-1}\times
   \nonumber
   \\
   &&\Big[1+(P_0^2+4p_\pi^2-M^2_{K^*})
   \Big(\int_0^\infty dr u(r)j_0(p_\pi r)e^{-M_{K^*}r}\Big)
   -4p^2_\pi\Big(\int_0^\infty dr u(r)(1+M_{K^*}r)\frac{j_1(p_\pi r)}{p_\pi r}
   e^{-M_{K^*}r}\Big)\Big]
   \label{e:RMB}
   \end{eqnarray}
    The $j_l(p_\pi r)$ are spherical Bessel functions. 
      Then Eq.~(\ref{e:gampipi}) reads
 
    \begin{eqnarray}
    \Gamma_{\pi\pi}=(\frac{2\pi\alpha^2_{\pi\pi}}{M^2_K}
    \frac{p_\pi}{M_K})\psi^2(0)R^2(p_\pi)=
    (v_{rel}\sigma_0)\psi^2(0)R^2(p_\pi)
  \label{e:gamfnl}
  \end{eqnarray}
  where  $v_{rel}=2p/M_K$ is the relative c.m. velocity
   of either kaon and $p_\pi=0.959\;M_K$ is now the pion momentum at threshold.
   For $R= 1$, Eq.~(\ref{e:gamfnl}) reduces to  the familiar
  ``wave function at contact'' approximation for the decay
  width \cite{TB85,MAS75,KLS04,RHL07} where only the
  probability per unit volume $\psi^2(0)$ of
   finding the pair at the origin characterizes the bound state.

   The isovector
   $K\bar K\to \pi^0\eta$
   decay width $\Gamma_{\pi^0\eta}$ is given by
   a similar expression with the following
   replacements in Eqs.~(\ref{e:sigzero}) to (\ref{e:gamfnl}):
   $p_\pi$ by  $p_{\pi\eta}=0.658\;M_K$, $P_0^2$ by
   $(E_\eta+\frac{1}{2}P_0)(E_{\pi^0}+\frac{1}{2}P_0)$, and $M^2_{K^*}$ by
   $\hat M^2_{K^*}=M^2_{K^*}+(E_\eta-\frac{1}{2}P_0)(E_{\pi^0}-\frac{1}{2}P_0)$.
   These modifications of the vertex in Eq.~(\ref{e:vertex}) reflect the revised total energy
   conservation condition, $P_0= E_{\pi^0}+E_\eta$ for an on--shell
   $\pi^0\eta$ pair in the final
   state with energies $E_{\pi^0}$ and $E_\eta$ and common c.m. momentum $p_{\pi\eta}$. Then
    $\alpha_{\pi\pi}$ is replaced by
 \begin{eqnarray}
 \alpha_{\pi\eta}=\frac{1}{4}\sqrt \frac{3}{2}\;\alpha_s 
 \Big[\frac{(E_\eta+\frac{1}{2}P_0)(E_{\pi^0}+\frac{1}{2}P_0)
 + p^2_{\pi\eta}}{\hat M^2_{K^*}+ p^2_{\pi\eta}} \Big]\approx
  1.079
  \label{e:apieta}
 \end{eqnarray}
  after removing the crossed contribution; $\psi(0)$  now refers
  to the isovector channel. 

   In Eq.~(\ref{e:gamfnl}) both $\psi(0)$ and $R$ are  functions of the
   binding energy; $R$ in particular only
   differs significantly from unity      
    when the ranges of the
    radial wave function $u(r)$ and the amplitude
    $M_{\pi\pi}({\bf p}_\pi, {\bf r})$  under the integrals
    in Eq.~(\ref{e:RMB}) are  commensurate.
    This is the case  for  the  analogous
    $M_{\gamma\gamma}({\bf p}_\gamma, {\bf r})$ that describes two--photon decay \cite{RHL07}.
    However for the loosely
     bound $K\bar K$ pairs in Tables~\ref{t:table1} and \ref{t:table2},
     $u(r)$ has a typical range $\sim M^{-1}_K$ 
     that is  about twice the range $ M^{-1}_{K^*}$ of the transition
     operator. Then  only the short
  range behavior of $u(r)$  is important for determining $R$. This behavior does not differ very  much from the
 small $r$ limit, 
 $u(r)=r$, when both integrals have the
  common
 value $(M^2_{K^*}+p^2_\pi)^{-1}$. This gives $R=1$, thereby reproducing the wave
 function at contact approximation.
  The same result 
 follows immediately from the momentum space representation of $M_B(p_\pi)$ 
 in the weak binding limit when  
 $\phi({\bf p})$ has a much shorter range than the transition operator in
 momentum space. Then
 the integral evaluates approximately as $M_{\pi\pi}({\bf p}_\pi, 0)\psi(0)$
 that leads directly back to Eq.~(\ref{e:MBfinal}) again  with $R=1$.

   The calculated widths are listed in Table~\ref{t:table1}. Since the cross section factors $ v_{rel}\sigma_I$ 
   for free annihilation are fixed by the coupling constants
   $\alpha_{\pi\pi}$, $\alpha_{\pi\eta}$, 
      the values of the
   decay widths given by Eq.~(\ref{e:gamfnl}) and its $I=1$ counterpart are controlled by the real $K\bar K$ potential that determines
    $\psi(0)$ and $R$ in each isospin channel. The cutoffs $\Lambda$
    have
    been fixed
    to reproduce a prescribed binding energy range. There are thus no free
    parameters.   The predicted widths  in Table I
    cover a range for $\Gamma_{\pi\pi}$ that agrees quite well with experiment within 
    the quoted error bars; the predicted values for $\Gamma_{\pi^0\eta}$ are too
    small.

  The actual values of  $R$ listed in Table~\ref{t:table2} for $I=0$ increase only slightly
  with binding energy from $1.034$ to $1.041$
 over the entire  range of $\epsilon_0$. The behavior for $I=1$ is similar.
 Thus the main variation in width in Table~\ref{t:table1} comes from the contact probability density $\psi^2(0)$ which
 increases approximately like $\sqrt \epsilon_I$; higher binding enhances the probability of contact
 at the origin. From a quark model perspective \cite{WIsgur82}
 this increased  contact probability facilitates strange quark annihilation and exchange
 to  produce the $(K\bar K)_0\to \pi\pi$ and 
 $(K\bar K)_1\to \pi^0\eta$ decays respectively. 

 \subsection*{Contributions from $s$-channel scalar meson exchange}
 The calculations of mass and widths just presented for the $K\bar K$ system
 ignore any $s$--channel contributions to the scattering amplitude
 $\hat \Gamma({\bf p}',{\bf p};P_0)$ in Eq.~(\ref{e:BSm}). However,
$K\bar K$ can also interact and annihilate via
 scalar meson exchange  in this channel.
One expects such  effects to be small because the exchanged particle is
off--shell. As an example we estimate their order of magnitude contribution for the case of
 $\sigma $ meson exchange, treated as a simple $q\bar q$ state.
 Then the relevant interactions can be read off from the
 $SU(3)$ linear sigma model (L$\sigma$M) Lagrangian \cite{MDS04,VD96,SCH71},
 \begin{eqnarray}
 {\cal L}_{cubic}= g_{\sigma\pi\pi}\sigma({\vec\pi}\cdot{\vec\pi})+g_{\sigma K \bar K}\sigma \bar K K
 +\cdots
 \label{e:cubicL}
 \end{eqnarray}
 The coupling constants are given by $ g_{\sigma\pi\pi}\approx M^2_\sigma/2f_\pi,
 g_{\sigma K \bar K}\approx M^2_\sigma/2f_K$
 in the (ideally mixed) chiral limit.
The $(f_\pi,f_K)\approx (92.2,110)$ MeV with $f_K/f_\pi\approx 1.19$ are the pion and kaon weak decay constants \cite{PDG08}.

\begin{figure}
\rotatebox{0}{\includegraphics{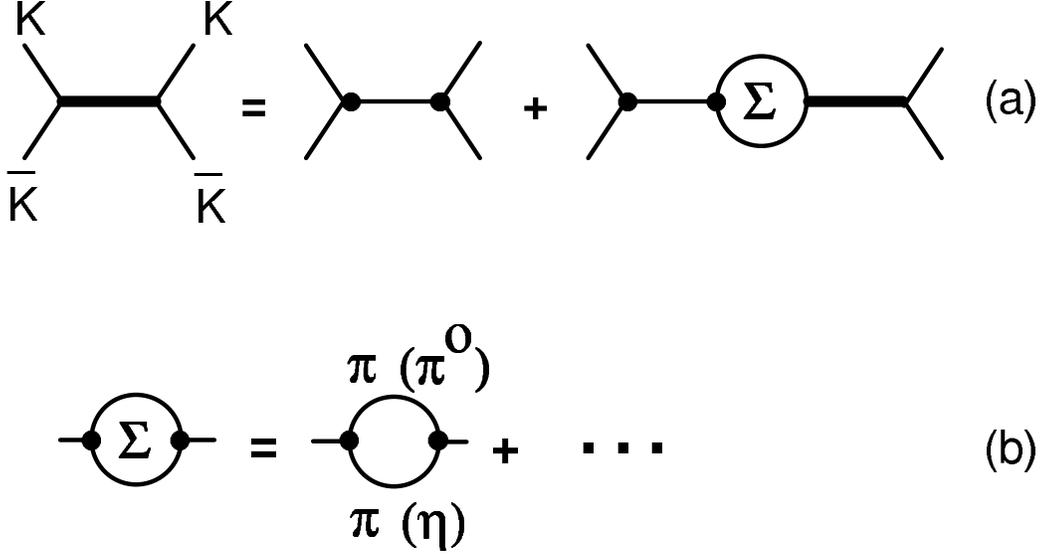}}
\caption{\label{f:fig3}
 (a) Meson exchange diagrams for $s$-channel contributions to the 
Bethe--Salpeter transition amplitude in Eq.~(\ref{e:BSm}). The thin
and thick solid lines indicate  bare versus dressed intermediate mesons
propagating between interaction vertices, shown as filled circles.
(b) The meson proper selfenergy $\Sigma$. Only the $\pi\pi$ or $\pi^0\eta$ polarization loop 
contributions to $\Sigma$ that include imaginary parts for dressing the
bare $\sigma$ or $a_0$ meson propagators at $s=P_0^2\approx 4M^2_K$ are shown.}
\end{figure}

 The relevant diagrams are shown in Fig.~\ref{f:fig3}. Translating these diagrams  
 one finds a $s$--channel   
 contribution to the BS equation scattering amplitude coming from $\sigma$  exchange
 as                                     
\begin{eqnarray}
\hat \Gamma_\sigma({\bf p}',{\bf p};P_0)= -(\sqrt 2g_{\sigma K\bar K})^2{\cal D}_\sigma(P_0^2))
\end{eqnarray}
for off--shell kaons with c.m. four--momentum squared  $s=P_0^2$, where
${\cal D}_\sigma(s)=[s-M^2_\sigma-\Sigma_\sigma(s)]^{-1}$ 
is the ``dressed'' scalar meson propagator \cite{LL74} of mass $M_\sigma$ and self--energy
 $\Sigma_\sigma(s)$.

The $s$-channel contribution to the additional binding energy and
decay width  for $I=0$ is then given in first order by 
\begin{eqnarray}
 \epsilon^{(\sigma)}_0+\frac{i}{2} \Gamma^{(\sigma)}_{\pi\pi}=\frac{1}{4M^2_K}\int \int \frac{d^3p'}{(2\pi)^3} \frac{d^3p}{(2\pi)^3}
  \phi({\bf p}')[\hat\Gamma_\sigma({\bf p}',{\bf p};P_0)]\phi({\bf p})= 
   - \frac{g_{\sigma K\bar K}^2}{2M^2_K}\psi^2(0){\cal D}_\sigma(P_0^2).
\label{e:Shiftwidth}
\end{eqnarray}
In particular,  
\begin{eqnarray}
\Gamma^{(\sigma)}_{\pi\pi}= 
\pi \frac{g_{\sigma K\bar K}^2}{M^2_K}\psi^2(0)\rho_\sigma(P_0^2), \quad
 \rho_\sigma(s)=-\frac{1}{\pi}Im{\cal D}_\sigma(s)
\label{e:Swidth}
\end{eqnarray}
 where $\rho_\sigma(s)$
 is the scalar spectral density associated with ${\cal D}_\sigma(s)$. 

 The self--energy $\Sigma_\sigma(s)$ receives contributions at one--loop level from
  tadpole and seagull diagrams in addition to  polarisation
  diagrams generated by Eq.~(\ref{e:cubicL}), plus  counter--terms. Of these,
  only the  polarization loop involving $\pi\pi$ for $I=0$ 
  (or $\pi\eta$ for  $I=1$) has
  an imaginary part for strong decay at the relevant
  four--momentum  transfer encountered in Eqs.~(\ref{e:Shiftwidth}) and (\ref{e:Swidth}).
  Cutting the loop in the standard way
  places  both pions on--shell and 
  gives the imaginary part  
  \begin{eqnarray}
 -Im\Sigma_\sigma(s)=  \frac{3g^2_{\sigma\pi\pi}}{8\pi}\Big({1-\frac{4M^2_\pi}{s}}\Big)^{1/2}
 \theta(s-4M^2_\pi)= M_\sigma\Gamma_\sigma(s),
 \label{e:ImSig}
 \end{eqnarray}
 The last equality defines
an ``off--shell'' width  $\Gamma_\sigma(s)$ that coincides with the physical
decay width for $\sigma\to \pi\pi$ at $s=M^2_\sigma$. 
Evaluating $\rho_\sigma(s)$  in Eq.~(\ref{e:Swidth}) at $s=P_0^2$,
one has
\begin{eqnarray}
 \Gamma^{(\sigma)}_{\pi\pi}= \frac{3p_\pi}{8\pi}\frac{\psi^2(0)}{M^3_K}
 \Big\{\frac{g^2_{\sigma\pi\pi}g^2_{\sigma K \bar K}}{(P_0^2-M^2_\sigma)^2
 +(Im\Sigma_\sigma(P_0^2))^2}
 \Big\}
 \label{e:delsig}
 \end{eqnarray}
after absorbing $Re \Sigma_\sigma(s)$ into a redefinition of the $\sigma$
mass $M^2_\sigma+(Re \Sigma_\sigma(P_0^2))^2\to M^2_\sigma$.
The c.m. momentum  $p_\pi$ refers to  outgoing
pions from the  $K\bar K\to \pi\pi$ decay as before. The binding energy contribution $\epsilon^{(\sigma)}_0$
follows from the real part of Eq.~(\ref{e:Shiftwidth}) as
\begin{eqnarray}
\epsilon^{(\sigma)}_0= (P_0^2-M^2_\sigma)[2Im\Sigma_\sigma(P^2_0)]^{-1}
\Gamma^{(\sigma)}_{\pi\pi}<0,
\label{e:deleps}
\end{eqnarray}
 i.e. the contribution to the real potential from the $s$-channel is
 repulsive (less binding) for $P_0\geq M_\sigma$.

We identify  $\sigma$ with the $f_0(600)$ scalar meson and  take its
mass and decay half--width parameters as $M_\sigma-\frac{i}{2}\Gamma_\sigma=
[(541\pm 39)-(252\pm 42)i]$ MeV directly from experiment \cite{BES04}.
By inverting Eq.~(\ref{e:ImSig}) for the $\sigma\to \pi\pi$ width
 the coupling constant is determined as $g_{\sigma\pi\pi}=1.63\pm 0.17$ GeV.
 Since this value is quite close to the chiral model  estimate of $g_{\sigma\pi\pi}
  = M_\sigma/2f_\pi \approx 1.48$ GeV, we assume that the L$\sigma$M
coupling constant ratio 
$(g_{\sigma\pi\pi})/(g_{\sigma K\bar K})
=f_K/f_\pi\approx 1.19$, also continues to hold to fix $g_{\sigma K\bar K}=1.37 $ GeV.
These parameters lead to $I=0$ contributions from the $s$-channel of 
\begin{eqnarray}
\epsilon^{(\sigma)}_0+\frac{i}{2}\Gamma^{(\sigma)}_{\pi\pi}=
 -8.92+3.93i,\quad -5.18+2.28i\quad{\rm and}\quad -2.92+1.29i\;{\rm MeV}
 \end{eqnarray}
 respectively for cutoffs  $\Lambda=415,\;420$ and $425$ MeV.

The exchange of other scalar mesons also has to be addressed. Since
both $f_0(980)$ and $a_0(980)$ are depicted as $K\bar K$ molecules in the
Weinstein--Isgur picture used here, the next available set of scalars to consider
would be the $0^+$ states $f_0(1370), K^*_0(1430), a_0(1450)$ and $f_0(1500)$ with masses above $1.3$ GeV.
 However a consistent theoretical description of these scalars is still
 under debate.
 If, as suggested in  \cite{VD96,MDS04,SCH71}, one places them in the same 
 $q\bar q$ scalar nonet and
implements a linear sigma model description, the  decay
widths of, for example,  $a_0(1450)$ into $K \bar K$ and 
 $\pi^0\eta$ come out far too large. This is in part due to the
large coupling constants generated by the model \cite{VD96}. On the other hand if one $\it{assumes}$
an interaction of the  L$\sigma$M form, but extracts 
the relevant coupling constants  $(g_{a_0\pi\eta},g_{a_0K\bar K})\approx (1.34,0.95)$ GeV
from the  $a_0(1450)$
branching ratios \cite{PDG08} into $\pi^0\eta\;(\sim 8\%)$
and $K\bar K\;(\sim 7\%)$, the $I=1$ contributions are
given by
\begin{eqnarray}
\epsilon^{(a_0)}_1+\frac{i}{2}\Gamma^{(a_0)}_{\pi^0\eta}=
2.16+0.085i,\quad 1.15+0.045i\quad {\rm and}\quad 0.59+0.023i\;{\rm MeV}
 \label{e:isovector}
\end{eqnarray}
for $s$--channel $a_0(1450)$ exchange. We omit the calculational details.

None of these contributions introduce any significant corrections into
 the $t$-channel values of Table~\ref{t:table1}. This is because
 the scalar spectral density is sampled at $s\approx 4M^2_K$ that pushes
 the exchanged meson  significantly off its mass--shell
to  weaken the $s$--channel $K\bar K$ interaction accordingly. In addition,
the small partial decay width $\sim 20$ MeV for $a_0(1450)\to \pi^0\eta$
relative to $\sigma\to \pi\pi$ serves to suppress the
$I=1$ contribution even further.

In closing this section we remark that the interaction vertices
described by the effective Lagrangian involving only meson degrees of freedom can
 also be
visualised at the constituent quark level. For instance the vertex
$K^++\bar K^{0\;*}\to \pi^+$, that contributes to $K^+K^-\to \pi^+\pi^-$ in Fig.~\ref{f:fig2}
translates schematically into
$[(u\bar s)(s\bar d) ]\to [u\bar d]$, the strange quarks $s\bar s$ having annhilated
(via  some unspecified gluon exchange interaction) leaving behind the $u$ and $\bar d$ quarks.

Moving beyond the effective meson
theory model then, this  suggests that in principle
the final two--meson decay channels can for example also be reached via
another route 
involving ``constituent gluons'' in intermediate states that give rise to hybrid meson structures
$[q\bar q g]$ (see \cite{KZ07}   
for a comprehensive review). Hybrids are generally expected \cite{Isgur85,IJG07}
to be heavier than $1.7-1.9$ GeV, thus lifting the threshold of such intermediate
states above $2$ GeV and making them less important than the intermediate meson states
which we have taken into account in the present context.
Moreover, the
hybrids' predicted decay properties  (for example no decays into two pseudoscalar
mesons, at least according to the gluon flux--tube model \cite{Close95} of their structure)
can only further reduce their impact here. Further discussion of such contributions lies
outside the scope of this article. We refer to the recent literature
\cite{KZ07,Isgur85,IJG07,Close95} for more details.

\section{$K\bar K$ Scattering}

 It is straightforward to show that the standard effective range expansion
 \cite{Bethe49} for the $K\bar K$ scattering phase shift $\delta_I(k)$
 at c.m. momentum $k=\sqrt{M_K E}$,
\begin{eqnarray}
 k\cot\delta_I(k)= -\frac{1}{a_I}+\frac{1}{2}r_I\;k^2+\cdots 
 \label{e:effrng}
\end{eqnarray}
 continues to
hold for the modified radial equation (\ref{e:radialeqn}) in terms of 
 the scattering length and effective range  $a_I$ and $r_I$.

In the presence
of annihilation both parameters pick up imaginary
contributions. In particular, in the limit $k\to 0$, the imaginary part of the
$I=0$ scattering length $a_0$ is related to the $K\bar K\to \pi\pi$
annihilation cross section $\tilde\sigma_0$ by
\begin{eqnarray}    
 -Im(a_0)&=&\frac{k}{4\pi}\tilde\sigma_0\approx
\frac{k}{4\pi}|f(0)|^{-2}\sigma_0= |f(0)|^{-2}\xi^{-1}_0
\nonumber
\\
\xi^{-1}_0&=& \frac{k}{4\pi}\sigma_0=\frac{\alpha^2_{\pi\pi}}{4M_K}
\frac{p_\pi}{M_K}=(1.881M_K)^{-1}.
\label{e:ima}
\end{eqnarray}
  The first step is exact. The next step follows after noting that $\tilde\sigma_0$ is  
 related to the free annihilation cross section $\sigma_0$ of
 Eq.~(\ref{e:sigzero}) by $\tilde\sigma_0\approx |f(0)|^{-2}\sigma_0$
 to a good approximation \cite {GW64}. 
  Here $f(k)$ is the Jost function \cite{GW64,VB49} 
 that determines the scattering matrix $S(k)=f(k)/f(-k)$. 
  The value $|f(0)|^{-2}$ at $k=0$ is the enhancement factor
  \cite{GW64} that gives the ratio of the probability of finding an interacting
  kaon pair at the origin to that when there is no interaction between them.

 If one parametrizes the Jost function $f(k)=|f(k)|\exp[i\delta_I(k)]$ as
 \begin{eqnarray}
 f(k)=\frac{k-ia}{k-ib}
 \end{eqnarray}
 the effective range expansion becomes
  exact \cite{VB49}, and scattering length and effective range are given 
in terms of the parameters ($a,b$). Considering $I=0$ again, one gets
\begin{eqnarray}
a_0=-(b-a)/ba,\quad r_0=2/(b-a).
\label{e:ab}
\end{eqnarray}

 Since the isoscalar
channel  supports a single
 $s$--wave bound state at complex binding energy
 $\tilde\epsilon_0=\epsilon_0+\frac{i}{2}\Gamma_{\pi\pi}$ one knows that
 the function $f(k)$ must
 have a zero \cite{VB49}  
 in the lower half of the complex $k$ plane at
 $k=ia=-i\sqrt{M_K\tilde\epsilon_0}$ which determines $a$. Combining Eqs.~(\ref{e:ima}) and
 (\ref{e:ab}) one then finds
  $b= \sqrt{\xi_0Im(-a)}$  for $b$ real. 
An identical procedure suffices for determining $(a,b)$ in the
isovector channel.

The scattering parameters for both channels are then given by
Eq.~(\ref{e:ab}). Taking the relevant values of $\tilde\epsilon_0$ and
 $\tilde\epsilon_1$ from Table I for $\Lambda=420$ MeV, one calculates  that
 \begin{eqnarray}
&&a_0=4.926-2.235i,\quad r_0=2.550-0.525i 
 \nonumber
 \\
&&a_1=8.246-2.576i,\quad r_1=2.971-0.242i 
\label{e:scatpar}
\end{eqnarray}
in units of $M^{-1}_K$.

\begin{figure}]
\rotatebox{0}{\includegraphics{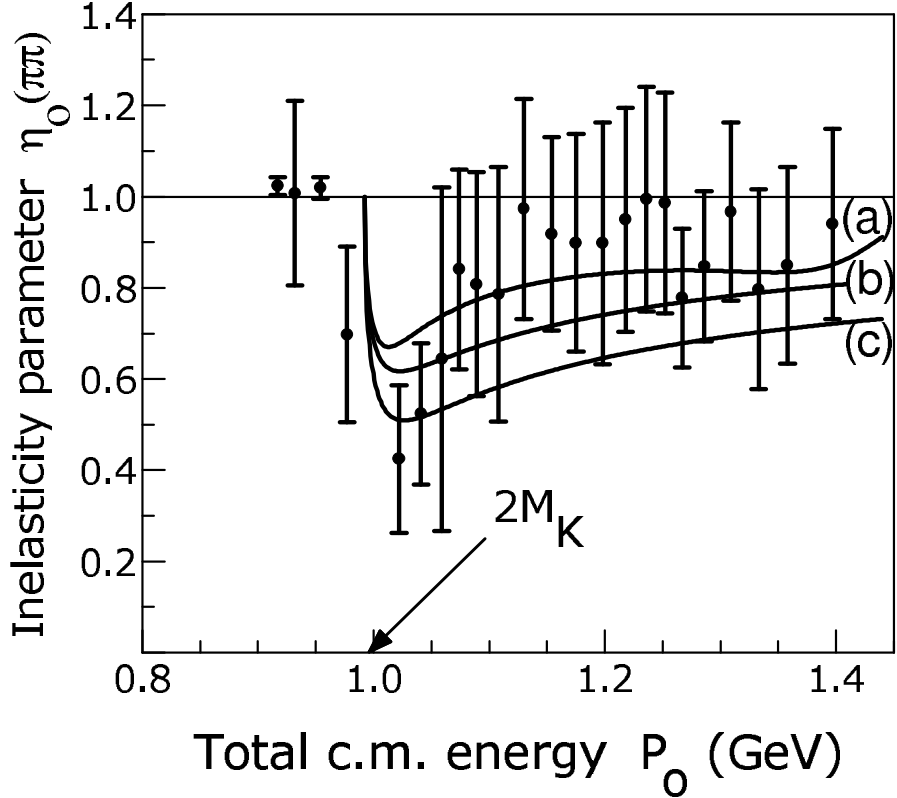}}
\caption{\label{f:fig4}
 The $\pi\pi$ inelasticity versus the total center of mass (c.m.) energy. The filled circles
 with error bars show values of $\eta_0(\pi\pi)$ taken from
 B. Hyams {\it et al.} \cite{BYM73}, while the upper curve (a) gives the 
  fit using the $K$--matrix from \cite{KPY06}. The lowest curve (c) shows
  the calculated inelasticity based on the effective range parameters of
  Eq.~(\ref{e:scatpar}). Curve (b) gives the result using the scattering
  parameters for the local potential in \cite{KLS04}.}

\end{figure}

Apart from one pioneering experimental attempt to determine the isoscalar scattering
length  by Wetzel ${et\;al.}$ \cite{WWE76} who find $a_0=[|(3.13\pm 0.30)|-(0.67\pm 0.07)i]M^{-1}_K$,
and a subsequent analysis  of later $\pi\pi$ data that infers \cite{RKLL95}
$a_0=(4.36-1.49i)\;M^{-1}_K$, 
there are as yet no other direct $K\bar K$ measurements with which to compare the
estimates in  Eq.~(\ref{e:scatpar}). However, to the extent that only coupling between the $\pi\pi$ and
$K\bar K$ channels is important, two channel unitarity shows \cite{JLP71}
that the $\pi\pi$ and $K\bar K$ isoscalar inelasticities $\eta_0(\pi\pi)=
\eta_0(K\bar K)=|S_0(k)|$ are equal at their common total c.m. energy
$P_0=2\sqrt{p^2+M_\pi^2}=2\sqrt{k^2+M_K^2}$ where $p$ and $k$ are the
pion and kaon  momenta. In Fig.~\ref{f:fig4} we compare $\eta_0(\pi\pi)$ as calculated
from $S_0(k)={\rm exp}[2i\delta_0(k)]$ in the effective range approximation, Eq.~(\ref{e:effrng}),
using the isoscalar parameters given above, with the inelasticities extracted
from $\pi\pi$ scattering in Ref.~\cite{BYM73}. The result using the isoscalar
scattering parameters $a_0=(4.281-2.398i)M^{-1}_K$,
$r_0=(1.169- 0.178i)M^{-1}_K$ obtained from
 the local potential in \cite{KLS04} is also
given for comparison.
A recent  $K$--matrix fit
based on a combination of various data sets
 is also shown, see \cite{KPY06} and further references cited therein. Given the
 very wide error bars on the data it is difficult to draw  definitive
 conclusions beyond remarking that the model calculations do
  reproduce the correct trend and order of magnitude of the $\pi\pi$ inelasticity,
  albeit somewhat too low.

   \section{kaonium}
                                                        
  We next summarize the  results obtained for the energy shifts and decay widths
  for the
  $K^+K^-$ (kaonium) atom  in the context of the present calculations.
  The properties of this system have already  been discussed in detail in \cite{KLS04}
  for local potentials.

  Pure Coulomb interactions bind kaonium at $-\frac{1}{2}\alpha^2\mu=-6.576$ keV
  in the lowest $1s$ state  where $\psi_{1s}(r)=\pi^{-1/2}(\mu\alpha )^{3/2}
  {\rm exp}(-\mu\alpha r)$; $\mu=\frac{1}{2}M_{K^\pm}$
  is the reduced mass  with $M_{K^\pm}\approx 494$ MeV, and
  $\alpha\approx 1/137$  the fine structure  constant.
  The kaonium $\to \pi\pi+\pi^0\eta$
  decay width follows from
  Eq.~(\ref{e:gamfnl})
after replacing  $\sigma_0$
by the total  $K^+K^-$ annihilation cross section $\sigma_p=(\sigma_0+\sigma_1)/2$
 and $\psi(0)$ by $\psi_{1s}(0)$.
  While
  $R$ can be calculated in closed form for the $1s$ Coulomb ground state of
  kaonium, 
  \begin{eqnarray}
   R_{1s}(p_\pi)=\Big[\frac{P_0^2+p^2_\pi}{M^2_{K^*}+p^2_\pi}\Big]^{-1}\Big[
   \frac{P_0^2+p^2_\pi-\mu\alpha(2M_{K^*}+3\mu\alpha)}{(M_{K^*}+\mu\alpha)^2+
    p^2_\pi}+4\frac{\mu\alpha}{p_\pi}\cot^{-1}
    \Big(\frac{(M_{K^*}+\mu\alpha)}{p_\pi}\Big)\Big]
  \end{eqnarray}
  this factor differs but little
 from unity, $R_{1s}(p_\pi)=1.0016$; the corresponding expression for $I=1$
 gives $R_{1s}(p_{\pi\eta})=1.0013$. This is to be expected.
 The Bohr radius of kaonium, $1/\mu\alpha =109$ fm, is about $500$ times  the transition
 amplitude range of $ M^{-1}_{K^*}\sim 0.2$ fm.
 Hence the wave function at contact is a very reliable approximation in this
 case too 
 and yields  a total $1s$ decay width of \footnote{
The analogous expressions quoted in \cite{YHPB06} for $M_B(p_\pi)$ and its
isovector counterpart do not take into account the delta function singularity
in the transition operator of Eq.~(\ref{e:transr}) in either channel, and
neglect crossing in the $\pi\pi$ channel.
As a result the kaonium decay
widths given there underestimate those in Eqs.~(\ref{e:coulwidth}) and (\ref{e:delE}) 
by an order of magnitude.}

 \begin{eqnarray}
 \Gamma_{1s}=\Gamma_{\pi\pi}+\Gamma_{\pi^0\eta}
 =\frac{1}{8}(\alpha_{\pi\pi}^2p_\pi+\alpha_{\pi\eta}^2p_{\pi\eta})\alpha^3
 =(51.3+18.4)\;{\rm eV}= 69.7\;{\rm eV},
 \label{e:coulwidth}
  \end{eqnarray}
or a lifetime of $\sim 10^{-17}$s.

 The effect of strong plus Coulomb interactions on the kaonium spectrum can be included
 in the standard way \cite{Bethe49} by equating the logarithmic derivative of the
 asymptotic form
 $u_p(r)\sim 1-\alpha_pr$ 
 of the zero energy  $K^+K^-$  scattering wave function outside the
 strong   interaction zone  with that of a pure incoming
 Coulomb wave function; $\alpha_p$ is given below.
 Physically  this  suggests that the binding of kaonium is still essentially provided by
 the long range Coulomb attraction.

 For bound states the Coulomb wave function is proportional to the
 Whittaker function \cite{RGN66}
 $f_c^{(-)}(k,r)\sim W_{i\eta,1/2}(2ikr)$ at complex wave number $k_\lambda=-i\lambda
 \mu\alpha$ that describes a decaying wave at $r\to \infty$.
 Here $\lambda$ is a yet to be determined
 eigenvalue that gives 
   the binding energies and total decay widths 
  $E_\lambda-\frac{1}{2}i\Gamma_\lambda=
  -\frac{1}{2}\lambda^2\mu \alpha^2$
  of the mesonic atom under the combined influence of strong and Coulomb
  interactions; $i\eta=\mu \alpha/ik_\lambda=1/\lambda$ is the Coulomb
  parameter for
  attractive interactions. The matching  condition at the
  Bethe ``Coulomb joining radius'' \cite{Bethe49} $d$ say,
  is then equivalent to
   the  Kudryavtsev--Popov  equation \cite{KP79}  for the complex eigenvalues $\lambda$
  that takes the simple form \cite{KLS04}
  \begin{eqnarray}
  (\alpha_p)_c=2\mu\alpha\Big[\psi(1-1/\lambda)+\frac{1}{2}\lambda
  +\ln\lambda+\gamma\Big]
  \label{e:KP}
  \end{eqnarray}
  for kaonium in the limit $\mu\alpha d<<1$, where   $\psi$ is the
  digamma function \cite{AS} and $\gamma=0.5771\cdots$ is Euler's
   constant. The $(\alpha_p)_c$ on the left  is the Coulomb corrected
   $K^+K^-$  inverse scattering length \cite{Bethe49},
 \begin{eqnarray}
   (\alpha_p)_c= \alpha_p-2\mu\alpha[\ln(2\mu\alpha d)+\gamma].
 \end{eqnarray}
    Here
    $\alpha_p=1/a_p$ 
    is the
    inverse of the $K^+K^-$ strong scattering length $a_p=(a_0+a_1)/2$ without Coulomb
    corrections if the isospin--breaking
    arising from the kaon mass difference
    $ \Delta=M_{K^0}-M_{K^\pm} \approx 4$ MeV \cite{PDG08} 
    is ignored. If not, the $K^+K^-\to K^0\bar K^0$ charge exchange
    channel is closed, and \cite{ADM70}
    \begin{eqnarray}                                                
    \alpha_p=\Big(\frac{1-k_0a_p}{a_p-k_0a_0a_1}\Big),
    \quad k_0=\sqrt{2M_{K^0}\Delta}\;.
    \label{e:AP}
     \end{eqnarray}
     We take
     the scattering lengths of good isospin from Eq.~(\ref{e:scatpar}),
     and calculate the Bethe
     joining radius  as $d=1.636M_K^{-1}$
   from the  regular zero energy
  scattering solution of Eq.~(\ref{e:radialeqn}). Then
  $\alpha_p= (6.873-3.387i)^{-1}M_K$ and
  $(\alpha_p)_c=(5.948-2.364i)^{-1}M_K$.
  Inserting this information on  the left hand side of
    Eq.~(\ref{e:KP})
   one finds $\lambda_{1s}=0.9579+0.0161i$, or
   \begin{eqnarray}                     
   E_{1s}-\frac{i}{2}\Gamma_{1s}=(-6.027-0.205i)\;{\rm keV}
   \label{e:1s}
   \end{eqnarray}
   for the lowest state.  Repeating the exercise for the remaining two values
   of the cutoff, the results for the energy shifts and decay widths for
   kaonium ground state
   can be summarized as follows,
   \begin{eqnarray}
   \Delta E_{1s}-\frac{i}{2}\Gamma_{1s}=(0.549_{-0.106}^{+0.059}-
   0.205_{-0.079}^{+0.100}i)\;{\rm keV},\quad \Lambda=420\pm 5\;{\rm MeV}
   \label{e:delE}
   \end{eqnarray}
   that illustrates the sensitivity to the choice of $\Lambda$. The corresponding
   lifetimes read $\tau=1.6^{-0.5}_{+1.00}\times 10^{-18}$s.

   Thus for $\Lambda=420$ MeV, apart from causing a repulsive level shift of
 $\Delta E_{1s}=0.549$ keV in the ground state energy of Coulombic kaonium, the strong
 interaction 
  also enhances the decay width
 considerably to $\Gamma_{1s}=0.410$ keV over that calculated in Eq.~(\ref{e:coulwidth}).
 There is thus strong mixing
 between the pure Coulomb state and the $f_0(980)$ molecular ground state
  that  produces a significant increase in decay width
 (and energy shift of the same order)
 that shortens the
 lifetime of the kaonium ground state to $\sim 2\times 10^{-18}$s.
 These results are in line with previous
 estimates based on a local potential description \cite{KLS04}.

\begin{figure}
\rotatebox{0}{\includegraphics{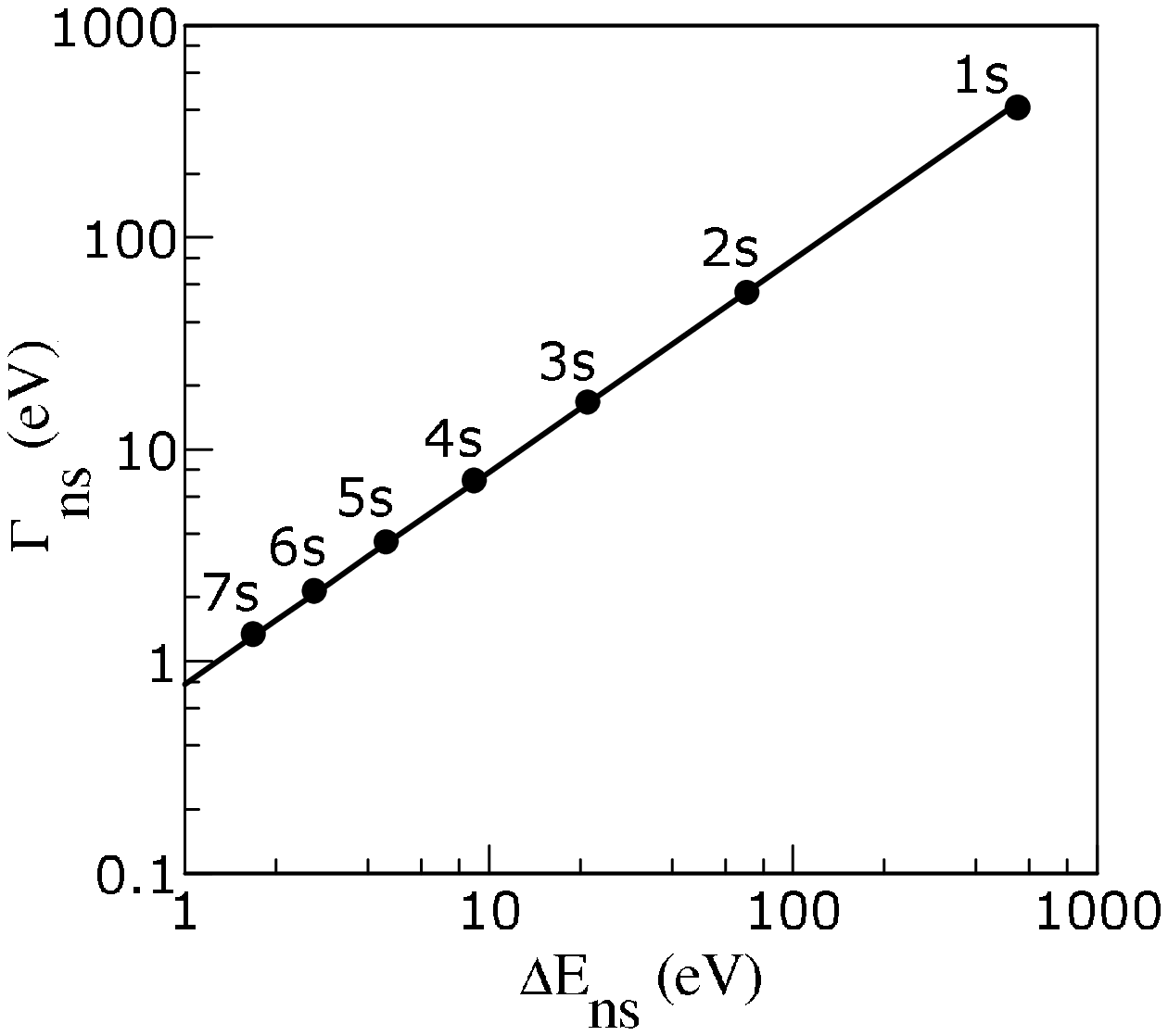}}
\caption{\label{f:fig5}
Calculated decay widths $\Gamma_{ns}$ versus the associated energy shifts
$\Delta E_{ns}$ 
for  kaonium plotted on a log--log scale. The straight line is given by
$\Gamma_{ns}=0.79\Delta E_{ns}$.}

 \end{figure}

 The strong mixing  feature $\Gamma_{1s}\simeq \Delta E_{1s}$ persists for the excited states of kaonium as well.
 This is illustrated for $\Lambda=420$ MeV in Fig.~\ref{f:fig5} which shows the 
 decay width versus the  energy shift for the kaonium spectrum.
 The ratio is seen to remain remarkably  constant over three orders
 of change in  magnitude in these quantities as one moves through the
 Balmer spectrum of kaonium. This can be understood from the
 Deser {\it et al.} \cite{DGBT54}                               
 perturbative solution of Eq.~(\ref{e:KP}) that is obtained by
 expanding the roots $\lambda^{-1}$ in
 powers of the corrected scattering length combination
 $\mu\alpha(a_p)_c=\mu\alpha(\alpha_p)_c^{-1}$ around their pure Coulomb field values at
  $\lambda_n^{-1}=n=1,2,3,\cdots$.  Then
 \begin{eqnarray}
 \Delta E_{ns}-\frac{i}{2}\Gamma_{ns}= \frac{2\mu^2\alpha^3}{n^3}(a_p)_c
 \Big[1+2\mu\alpha(a_p)_c\Big(\psi(n)-\psi(1)+\frac{1}{2n}-\ln n\Big)+\cdots\Big]
 \end{eqnarray}
 that has a very similar structure to the analogous chiral perturbation theory result
 for kaonic hydrogen \cite{UGM04}. Working to lowest order, one retrieves the Deser formula 
 \begin{eqnarray}
 \Delta E_{ns}-\frac{i}{2}\Gamma_{ns}\approx \frac{2\mu^2\alpha^3}{n^3}
 (a_p)_c =(0.575-0.228i)\;{\rm keV}, \quad n=1,
  \end{eqnarray}
 which gives  $n=1$ shift and width values that are $5$ and $11\%$ too large in comparison with Eq.~(\ref{e:delE}).
 The relation
  $\Gamma_{ns}= -2Im(a_p)_c/Re(a_p)_c\Delta E_{ns}= 0.79\Delta E_{ns}$
 that follows from the latter result  gives the straight line in Fig.~\ref{f:fig5}.

 The  underlying  reason for the  mixing can be qualitatively understood by
 temporarily ignoring annihilation   so that $\alpha_p$ is real and positive. Then one sees 
 from the asymptotic form of $u_p(r)$ that the
 kaonium ground  and excited state wave functions of Fig.~\ref{f:fig5} all develop a common
 additional node at $r=1/\alpha_p$ that renders them orthogonal to the ground state wave function of the bound
 $K\bar K$ pair. They are consequently all excited states
 in the combined strong plus Coulomb potentials, that suffer level
 repulsion with the kaon-antikaon ground state.

 \section{Summary and conclusions}
 It is shown that the $K\bar K$ momentum--dependent interaction
 due to vector meson exchange can be incorporated without further approximation
 into  a non--relativistic Bethe--Salpeter  wave equation in coordinate space
 containing, in addition to a local potential, a spatially dependent
 effective kaon mass embedded in a fully symmetized
 kinetic energy operator.

A numerical analysis of the possible bound and
scattering states of this modified equation is presented  as a model
for the properties of a molecular kaon--antikaon description of the light
scalar mesons $f_0(980)$ and $a_0(980)$. We show that
for a restricted range of  cutoffs $\Lambda$, the only adjustable
parameter in the calculations,
one obtains bound state masses  for both scalar mesons  in reasonable accord
with experiment.   
The accompanying
$\pi\pi$ and $\pi^0\eta$
decay widths exhibit a substantial variability with $\Lambda$ since they are sensitive to the contact probability density as
determined by the  ground state eigenfunction.
The results can be displayed in round numbers as follows
\begin{eqnarray}
M_0-\frac{i}{2}\Gamma_{\pi\pi}=[(988\;{\rm to}\;971)-\frac{i}{2}(36\;{\rm to}\;108)]\;{\rm MeV};\quad
M_1-\frac{i}{2}\Gamma_{\pi^0\eta}=[(989\;{\rm to}\;976)-\frac{i}{2}(7\;{\rm to}\;27)]\;{\rm MeV}
\label{e:summary}
\end{eqnarray}
as the cutoff $\Lambda$ increases from $415$ to $425$ MeV. The predicted range for $\Gamma_{\pi\pi}$ overlaps
nicely with the experimental spread of values $(40-100)$ MeV; the decay widths $\Gamma_{\pi^0\eta}$ for
molecular $a_0(980)$ 
are too small. We also demonstrated at the end of Section III that the associated $s$--channel contributions
to the  binding energy and decay widths in both isospin channels are unimportant.

These calculations treat the decay vertices as point--like.
Introducing a form factor like that of Eq.~(\ref{e:ff})
replaces $R\sim 1$ by $ R_{\Lambda}\sim 0.34$ in  the transition
amplitude of Eq.~(\ref{e:MBfinal}) if we assume the same cutoffs again as used for the potentials,
see Appendix Eq.~(\ref{e:Feff}).
Then one obtains an order of magnitude
reduction in widths down to $\Gamma_{\pi\pi}\simeq (4-12)$
and $\Gamma_{\pi^0\eta}\simeq (1-3)$ MeV from those given in Eq.~(\ref{e:summary}).
This agrees qualitatively
 with the calculations in \cite{XHG07} that also find small decay widths of
this order from a numerical study of the
 relativistic  Bethe--Salpeter
integral equation for the $K\bar K$ system when form factors are included.
They in addition find that bound states  only occur for cutoffs $\Lambda$ lying in a
series of narrow windows  similar to the one illustrated in Fig.~\ref{f:fig1}
for the non--relativistic case.

The low energy  parameters for $K\bar K$ scattering
in the presence of  annihilation  have also been
calculated in both isospin channels.
In particular this information allows one to construct
the $K\bar K$ isoscalar inelasticity as a function of the total energy
in order to compare with the measured $\pi\pi$ inelasticity which it
equals in  the two--channel problem considered here. The qualitative agreement
with experiment is satisfactory.

Finally, the effect of strong interactions on the binding energy and decay of
the $K^+K^-$ mesonic atom are briefly discussed in
the context of  momentum--dependent potential model. The energy shifts and decay
widths so obtained are again not very different from those using the local
potential approximation, and give a  lifetime of $\sim 2\times 10^{-18}$s for kaonium.

Overall, one can conclude that both the local  and non--local
potential description for
 the interaction  of a $ K\bar K$ pair lead to very similar results that are only dependent
on the  cutoff required
 in the  form factor of the potential for obtaining bound states. However the width
predictions of either of these, or other \cite{XHG07}, calculations  remain
 ambiguous, depending as they do on whether the vertices in the transition amplitudes
are regarded as point--like or extended.
The calculation  from first principles of
such hadronic size effects lies beyond the scope of the present work.
Without a better understanding of
this latter aspect, one cannot decide on the basis of width predictions alone
 between the molecular versus  elementary particle  interpretations
of the dynamics governing the structure of the light  scalar mesons.
 What has been shown here, however, is that the molecular model for
 $f_0(980)$ and $a_0(980)$
 with either local or non--local meson exchange interaction potentials  does give
 reasonable mass and width predictions for
 the isoscalar channel in particular, assuming  point--like vertices in the decay amplitudes.
In this regard, observation of the energy shift and decay width in kaonium, which in combination directly probe
the strong $K^+K^-$ scattering length, would be a particularly important  source of experimental information for further progress
towards a better understanding of these interesting systems.

\section{Acknowledgments}

     This research was  supported
    by a grant from the Ernest Oppenheimer Memorial Trust, which is gratefully
    acknowledged. I would also like to thank David Bugg for bringing
    Ref.~\cite{BES04} to my attention, and Veljko  Dmitrasinovi$\acute{c}$ for
    constructive correspondence.

\appendix*
\section{}
\subsection*{Decay widths with form factors}
 In order to include a form factor
in the transition operator $M_{\pi\pi}({\bf p}_\pi,{\bf r})$ of Eq.~(\ref{e:transr})
 we  replace the pure Yukawa
potential in that equation
by $U_i(r)$ from Eq.~(\ref{e:basicpot})  with $M_i=M_{K^*}$ and
$c_i=M_{K^*}/\sqrt{2}\Lambda$.
Then the revised transition amplitude splits into two parts,
\begin{eqnarray}
 M_{\Lambda,\pi\pi}({\bf p}_\pi,{\bf r})
=&&-2\sqrt 3 g^2_{\pi K K^*}e^{-i{\bf p}_\pi\cdot{\bf r}}\times
\nonumber\\
&&\Big[\delta^3_\Lambda ({\bf r})
-\{P^2_0+4p^2_\pi+4i{\bf p}_\pi\cdot{\bf \nabla}\}
U_\Lambda(r)
+\{P^2_0+4p^2_\pi-M^2_{K^*}+
4i{\bf p}_\pi\cdot{\bf \nabla}\}\frac{e^{-M_{K^*}r}}{4\pi r}\Big]
\label{e:MLAMpipi}
\end{eqnarray}
after using
\begin{eqnarray}
-\nabla^2U_i(r)=-M_{K^*}^2\frac{e^{-M_{K^*}r}}{4\pi r}+\delta^3_\Lambda({\bf r})
\end{eqnarray}
where
\begin{eqnarray}
 U_\Lambda(r)= \frac{e^{-\sqrt 2 \Lambda r}}{4\pi r}\sum_{n=0}^3 C^{(0)}_n(\sqrt 2\Lambda r)^n
\end{eqnarray}
and
\begin{eqnarray}
 && \delta^3_\Lambda({\bf r})= (\sqrt{2}\Lambda)^2
   \frac{e^{-\sqrt 2 \Lambda r}}{4\pi r}\sum^3_{n=0} C^{(2)}_{n}
   (\sqrt 2 \Lambda r)^n.
\label{e:spreaddel}
\end{eqnarray}
The coefficients  $C^{(2)}_n$ are linear combinations of the
$C^{(0)}_n$: $C^{(2)}_0=(C^{(0)}_0-2C^{(0)}_1+2C^{(0)}_2)$,
$C^{(2)}_1=(C^{(0)}_1-4C^{(0)}_2+6C^{(0)}_3)$, $C^{(2)}_2 = (C^{(0)}_2-6C^{(0)}_3)$ and
$C^{(2)}_3 = C^{(0)}_3$.
Since
$\delta^3_\Lambda ({\bf r})\to 0\;{\rm or}\;\infty$ depending on the order of the
limits $\Lambda \to \infty,\;r\to 0$, while
\begin{eqnarray}
\int d^3r \delta^3_\Lambda ({\bf r})=\sum_{n=0}^3C^{(2)}_n\Gamma(n+2)=1
\end{eqnarray}
independent of $\Lambda$, the function defined by
Eq.~(\ref{e:spreaddel}) is a representation of a spread--out delta function for
 finite $\Lambda$. Clearly 
 $M_{\Lambda,\pi\pi}({\bf p}_\pi,{\bf r})$ reverts back to
$M_{\pi\pi}({\bf p}_\pi,{\bf r})$ when $\Lambda\to\infty$. 
For $\Lambda\neq \infty$ Eq.~(\ref{e:MLAMpipi}) shows that the factor $R(p_\pi)$ in the bound to free transition amplitude of
Eq.~(\ref{e:MBfinal}) is replaced by 
\begin{eqnarray}
&&R_{\Lambda}(p_\pi)= \Big[\frac{P^2_0+p^2_\pi}{M^2_{K^*}+p^2_\pi}\Big]^{-1} 
\Big[\Big((\sqrt 2\Lambda)^2 \sum_{n=0}^3C^{(2)}_n
-(P^2_0+4p^2_\pi) \sum_{n=0}^3C^{(0)}_n\Big)
\Big(\int_0^\infty dr u(r)j_0(p_\pi r)
 (\sqrt 2\Lambda r)^n e^{-\sqrt 2 \Lambda r}\Big)
\nonumber\\
&&+4p^2_\pi\Big(\int_0^\infty dr u(r)[(1+\sqrt 2\Lambda r)
+\sum_{n=2}^4C^{(1)}_n(\sqrt 2 \Lambda r)^n]\frac{j_1(p_\pi r)}{p_\pi r}e^{-\sqrt 2 \Lambda r}\Big)
\nonumber\\
&&+(P^2_0+4p^2_\pi-M^2_{K^*})\Big(\int_0^\infty dr u(r)j_0(p_\pi r)
  e^{-M_{K^*} r}\Big)-4p^2_\pi\Big(\int_0^\infty dr u(r)(1+M_{K^*}r)\frac{j_1(p_\pi r)}{p_\pi r}
e^{-M_{K^*} r}\Big)\Big]
\nonumber\\
\label{e:RMBLAM}
\end{eqnarray}
with $C^{(1)}_2 =(C^{(0)}_1-C^{(0)}_2)$, $C^{(1)}_3 =(C^{(0)}_2-2C^{(0)}_3)$ and
$C^{(1)}_4 =C^{(0)}_3$.
Setting $u(r)\to r$ again that characterizes the weak binding limit
one confirms by direct calculation that now 
\begin{eqnarray}
R_{\Lambda}(p_\pi)=\Big(\frac{2\Lambda^2-M^2_{K^*}}{2\Lambda^2+p^2_\pi}\Big)^4=F_i^2(p^2_\pi)
\label{e:Feff}
\end{eqnarray}
instead of unity, where $F_i(p^2_\pi)$ is the form factor given by Eq.~(\ref{e:ff})
evaluated at $M_i=M_{K^*}$ and ${\bf q}^2=p^2_\pi$. For $\Lambda=420$ MeV
the actual values are $R_{\Lambda} =0.39$ versus $F_i^2=0.34$ so the wave function at contact approximation
continues to hold as well. Thus, depending on the value of the cutoff chosen, including a form factor in the transition amplitudes
 can reduce both the $\pi\pi$ and $\pi^0\eta$ widths in
 Table~\ref{t:table1} by an order of magnitude.

\newpage

\begin{table}[ht]
\caption{\label{t:table1}
Calculated values (in MeV) of the isoscalar ($\epsilon_0$) and
isovector ($\epsilon_1$)
 binding energies and resulting masses $M_I=2M_K-\epsilon_I$ and annihilation widths
for $f_0(980)$ and $a_0(980)$ for varying cutoff $\Lambda$  as given by the non--local meson exchange
potential model. Theoretical and experimental
values from various  sources have also been listed.}

\begin{ruledtabular}
\begin{tabular}{ccccc}
$\Lambda$ 
&
$\epsilon_0$ 
& $M_{0}-\frac{i}{2}\Gamma_{\pi\pi}$ 
& $\epsilon_1$  
& $M_{1}-\frac{i}{2}\Gamma_{\pi^0\eta}$ \\
\colrule
\\
  $415$  &   $21.3$  & $971-54\;i$ & $15.5$ & $976-13.7\;i$\\
\\
  $420$ &  $10.6$ & $981-33\;i$ & $7.55$ & $984-7.30\;i$\\
\\
  $425$ &  $4.03$ & $988-18\;i$ & $2.56$ & $989-3.74\;i$\\
\\
Local potential, Ref.~\cite{KLS04} &      $18.6$& $981-25\;i$ & unbound
&  --  \\
\\
Linear $\sigma$ model, Ref.~\cite{MDS04} &   $ - $& $940-27\;i$ & --
&  $983.4-69\;i$  \\
 \\
  Fermilab E 791 Collaboration \cite{E791} &-- &
   $(975\pm 3)-(22\pm 2)\;i$ &--&-- \\
\\
 KLOE Collaboration \cite{KLOE02} &--& --
 &-- & $(984.8\pm 1.2)-61\;i$\\

\\
 OBELIX Collaboration \cite{OBELIX03} &--& --
 &{--} & $(998\pm 10)-23\;i$\\

\\
 Particle Data Group \cite{PDG08} &--& $(980\pm 10)-(20\; {\rm to}\; 50)\;i$
 &-- & $(984.7\pm 1.2)-(25\; {\rm to}\; 50)\;i$\\

 \end{tabular}
\end{ruledtabular}
\end{table}

\begin{table}[ht]
\caption{\label{t:table2}
Bound state properties of the isoscalar potential: depth $V_0$, Eq.~(\ref{e:vezero}),
effective mass ratio, Eq.~(\ref{e:effmass}),
 binding energy $\epsilon_0$, wave function at contact amplitude $\psi(0)$,
 and  the factor $R$ of Eq.~(\ref{e:RMB}).}

\begin{ruledtabular}
\begin{tabular}{cccccc}
$\Lambda$ (MeV)
&
$V_0(0)$ (GeV) 
&
$M^*_K(0)/M_K$
&
$\epsilon_0$ (MeV)
&
$\psi(0)/M_K^{3/2}$
&
$R$ \\

\colrule
\\
  $415$  &  $-0.23$ & $2.18$ & $21.3$ & $0.126$ & $1.041$\\ 
\\
  $420$ & $-0.21$ & $1.93$ & $10.6$ & $0.096$ & $1.038$\\
\\
  $425$ & $-0.19$ & $1.75$ & $4.03$ & $0.072$ & $1.034$\\ 
 \end{tabular}
\end{ruledtabular}
\end{table}

 \newpage
\clearpage  

\begin{thebibliography}{17}



\bibitem{WIsgur82}
J. Weinstein and N. Isgur, Phys. Rev. Lett. {\bf 48}  659 (1982); Phys. Rev.
D {\bf 27}, 588 (1983); D {\bf 41}  2236 (1990).

\bibitem{TB85}

T. Barnes, Phys. Lett. {\bf 165 B}  434 (1985).

\bibitem{SU390}
D. Lohse, J. W. Durso, K. Holinde and J. Speth, Nucl. Phys. {\bf A516},
513 (1990); G. Jan$\beta$en, B. C. Pearce, K. Holinde and J. Speth, Phys. Rev.
 D {\bf 52}, 2690 (1995).



\bibitem{JAO03}

J. A. Oller, Nucl. Phys. {\bf A714}, 161 (2003).

\bibitem{KLS04}
S. Krewald, R. H. Lemmer and F. P. Sassen, Phys. Rev. D {\bf 69},
 016003 (2004).

\bibitem{TBR08}
 T. Branz, T. Gutsche and V.E. Lyubovitshij, Eur. Phys. J. A {\bf 37}, 303 (2008).

\bibitem{LM00}

L. Montanet, Nucl. Phys. B (Proc. Suppl.) {\bf 86}  381 (2000); V.V. Anisovich
{\it et al.}, Phys. Lett. {\bf B480}  19 (2000).

\bibitem{NNA89}

N. N. Achasov and V. N. Ivanchenko, Nucl. Phys.  {\bf B315}  465 (1989).

\bibitem{RKNI87}

R. Kokoski and N. Isgur, Phys. Rev. D {\bf 35}  907 (1987).

\bibitem{RT02}

R. H. Lemmer and R. Tegen, Phys. Rev. C {\bf 66}  065202 (2002).
 
\bibitem{PDG08}
C. Amsler {\it et al.}, (Particle Data Group),
 Phys. Lett. {\bf B667}, 1 (2008).



\bibitem{MDS04}
M. D. Scadron, G. Rupp, F.Kleefeld, and E. van Beveren, Phys. Rev. D {\bf 69}, 014010 (2004); Erratum-ibid.
D {\bf 69}, 059901 (2004).

\bibitem{VD96}
V. Dmitrasinovi$\acute{c}$, Phys. Rev. C {\bf 53}, 1383 (1996).




\bibitem{PEN07}

M. R. Pennington, Prog. Theor. Phys. Suppl. {\bf 168}, 143 (2007)
[arXiv:hep-ph/0703256v1].


\bibitem{LL74}
See, for example, L. D. Landau and E. M. Lifshitz, {\it Relativistic quantum
theory}, Course of Theoretical Physics Vol. 4 (Pergammon, Oxford, 1974).



\bibitem{WLFS01}
 W.L. Lucha, K. Maung Maung and F. F. Sch\"{o}berl, Phys. Rev. D {\bf 64}, 036007 (2001).



\bibitem{YHPB06}
Y.--J. Zhang, H.--C. Chiang, P.--N. Shen and B.--S. Zou, Phys. Rev. D {\bf 74},
014013 (2006).



\bibitem{DESWART63}
J. J. de Swart, Rev. Mod. Phys. {\bf 35}  916 (1963).

\bibitem{KSRF}
K. Kawarabayashi and M. Susuki, Phys. Rev. Lett, {\bf 16}, 255 (1966);
X. Riazuddin  and X. Fayyazuddin, Phys. Rev. {\bf 147}, 1071 (1966).

\bibitem{BUGG04}
F. Q. Wu, B. S. Zou, L. Li and D. V. Bugg, Nucl. Phys. {\bf A735}, 111 (2004).


\bibitem{WFRL57}
W. E. Frahn and R. H. Lemmer, Nuovo. Cim. {\bf 5}, 523, 1564 (1957); $ibid$
{\bf 6}, 1221 (1957).

\bibitem{PFRL96}
For recent reviews and applications see  P. Fr\"{o}brich and R. Lipperheide,
{\it Theory of Nuclear Reactions}, (Oxford: Clarendon Press, New York: Oxford
University Press, 1996);
A. Lovell and K. Amos, Phys. Rev. C {\bf 62}, 064614 (2000).

\bibitem{E791}

Fermilab E791 Collaboration, E. M. Aitala {\it et al.}, Phys. Rev. Lett.
{\bf 86}  765 (2001).

\bibitem{KLOE02}

KLOE Collaboration, A. Aloisio, {\it et al.}, Phys. Lett. {\bf B536}, 209 (2002).

\bibitem{OBELIX03}

OBELIX Collaboration, M. Bargiotti, {\it et al.}, Eur. Phys. J.
C {\bf 26}, 371 (2003).


\bibitem{RHL07}

R. H. Lemmer, Phys. Lett. {\bf B650}, 152 (2007).

\bibitem{HANR07}

C. Hanhart, Yu. S. Kalashnikova, A. E. Kudryavtsev, and A. V. Nefediev,
Phys. Rev. D {\bf 75}, 074015 (2007).

\bibitem{MAS75}

J. A. Wheeler, Ann. NY Acad. Sci. {\bf 48}  219 (1946);
M. A. Stroscio, Phys. Rep. C {\bf  22},  215 (1975).

\bibitem{SCH71}
J. Schechter and Y. Ueda, Phys. Rev. D {\bf 3}, 2874 (1971).

\bibitem{BES04}
BES Collaboration, M. Ablikim, {\it et al.}, Phys. Lett. {\bf B598}, 149 (2004).

\bibitem{KZ07}
E. Klempt and A. Zaitsev, Phys. Rep. {\bf 454}, 1 (2007).

\bibitem{Isgur85}
 N. Isgur and J. E. Paton, Phys. Rev. D {\bf 31}, 2910 (1985).


\bibitem{IJG07}
I. J. General, S. R. Cotanch, and F. J. Llanes--Estrada, Eur. Phys. J. {\bf  C 51},
347 (2007).

\bibitem{Close95}
F. E. Close and P. R. Page, Phys. Rev. D {\bf 52}, 1706 (1995).


\bibitem{Bethe49}
H. A. Bethe, Phys. Rev. {\bf 76}, 38 (1949); J. D. Jackson and J. M. Blatt,
Rev. Mod. Phys. {\bf 22}, 77 (1950).


\bibitem{GW64}

See, for example, M. L. Goldberger and K. M. Watson, {\it Collision Theory},
(John Wiley and Sons, New York, 1964), chapter 9.


\bibitem{VB49}
V. Bargmann, Rev. Mod. Phys. {\bf 21},  488 (1949);
R. G. Newton, J. Math. Phys. {\bf 1}, 319 (1960).

\bibitem{WWE76}
W. W. Wetzel {\it et al.}, Nucl. Phys. {\bf B115}, 208 (1976).

\bibitem{RKLL95}

R. Kami\'{n}ski and L. Le\'{s}niak, Phys. Rev. C {\bf 51}, 2264 (1995).


\bibitem{JLP71}
J. L. Petersen, Phys. Rep. {\bf 2C}, 155 (1971).


\bibitem{BYM73}

B. Hyams {\it et al.}, Nucl. Phys. {\bf B64}, 134 (1973).

\bibitem{KPY06}

R. Kami\'{n}ski, J. R.Pelaez, and F. J. Yndur\'{a}in, Phys. Rev.
D {\bf 74}, 014001 (2006); Phys. Rev. D {\bf 74},  079903(E) (2006).

\bibitem{RGN66}
See, for example, R. G.Newton, {\it Scattering Theory of Waves and Particles}
(McGraw--Hill, New York, 1966).

\bibitem{KP79}

A.E. Kudryavtsev and V. S. Popov, JETP Lett. {\bf 29}, 280 (1979); V. S. Popov,
A.E. Kudryavtsev, and V. D. Mur, Sov. Phys. JETP {\bf 50}, 865 (1979).

\bibitem{AS}
 M. Abramowitz and I. A. Stegun, Editors, {\it Handbook of Mathematical
 Functions} (Dover Publications, Inc., New York, 1965).


\bibitem {ADM70}
A. D. Martin and G. G Ross, Nucl. Phys. {\bf B16}, 479 (1970).


\bibitem{DGBT54}

S. Deser, M. L. Goldberger, K.Baumann, and W. Thirring, Phys. Rev. {\bf 96},
774 (1954).

\bibitem{UGM04}
Ulf--G. Mei$\beta$ner, U. Raha and A. Rusetsky, Eur. Phys. J. C {\bf 35}, 349
(2004).

\bibitem{XHG07}
X--H. Guo and X--H. Wu, Phys. Rev. D {\bf 76}, 056004 (2007).


 \end{thebibliography}
\end{document}